\documentclass[epj]{svjour}
\usepackage{latexsym}
\usepackage{amsfonts}
\begin{document}
\title{On the rational solutions of the $\widehat{su}(2)_k\,$
Knizhnik-Zamolodchikov equation}
\author{Ludmil Hadjiivanov
\and
Todor Popov
}                     
%
%
\institute{Theoretical Physics Division,
Institute for Nuclear Research and Nuclear Energy,\\
Tsarigradsko Chaussee 72, BG-1784 Sofia, Bulgaria}
%
\date{Received: date / Revised version: date}
%
\abstract{
We present some new results on the rational solutions
of the Knizhnik-Zamolodchikov (KZ) equation for the four-point
conformal blocks of isospin $I\,$ primary fields in the $SU(2)_k\,$
Wess-Zumino-Novikov-Witten (WZNW) model. The rational solutions
corresponding to integrable representations of the affine algebra
$\widehat{su}(2)_k\,$ have been classified in \cite{MST}, \cite{ST};
provided that the conformal dimension is an integer,
they are in one-to-one correspondence
with the local extensions of the chiral algebra. Here we give another
description of these solutions as specific braid-invariant combinations
of the so called regular basis introduced in \cite{STH} and
display a new series of rational solutions for isospins $I = k+1\,,\ k \in \Bbb N\,$
corresponding to non-integrable representations of $\widehat{su}(2)_k\,.$
}

\maketitle


\def\be{\begin{equation}}
\def\ee{\end{equation}}
\def\ba{\begin{eqnarray}}
\def\ea{\end{eqnarray}}
\def\lb{\label}
\def\a{\alpha}
\def\b{\beta}
\def\g{\gamma}
\def\d{\delta}
\def\i{\eta}
\def\e{\varepsilon}
\def\x{\xi}
\def\l{\lambda}
\def\s{\sigma}
\def\t{\tau}
\def\r{\rho}
\def\v{\varphi}
\def\D{\Delta}
\def\G{\Gamma}
\def\L{\Lambda}
\def\P{\Phi}
\def\E{{\cal E}}
\def\cC{\cal C}
\def\fp{{\frak p}}
\def\C{\Bbb C}
\def\Z{\Bbb Z}
\def\F{\Bbb F}
\def\R{\hat{R}}
\def\Rp{\hat{R}(p)}
\def\uz{\underline z}


\section{Introduction}
\label{intro}
The $2D$ nonlinear $\s$-model with a suitably normalized WZ term,
known as WZNW model \cite{W}, is a conformally invariant
(and therefore integrable) field theory with a huge internal symmetry,
beautiful geometric structure at the classical level and rich algebraic content
in the quantized case. The model describes a closed (respectively, open, for the
so called boundary model) string moving freely on a Lie group
manifold $G\,.$ After choosing
the group, the only parameter left which fixes the theory is a positive integer $k\,$
playing the role of WZ term coupling constant. Here
we will only consider the case of the compact group $G=SU(2)\,.$

To solve the model, one can use different approaches
in both the classical and the quantum cases.
In the {\em axiomatic} approach to the quantized model
one constructs the space of states
as a direct sum of superselection sectors (tensor products of integrable
representations of the corresponding left and right current algebras, both of which
appear to be affine algebras of the type ${\hat{\cal G}}_k\,$ where ${\cal G}\,$
is the Lie algebra of $G\,$ and $k\,$ is the level).
Each sector is generated from the vacuum by a primary field.
The interplay of affine and conformal invariance leads to linear
systems of partial differential equations
for the correlation functions of primary fields
(conformal blocks), one for each set of chiral
variables. These are the famous KZ equations
\cite{KZ}, \cite{T} determining, in principle, the chiral structure of the
theory; the correlation functions of the
$2D$ theory are recovered by combining the left and the right conformal
blocks (which are, typically, multivalued) in such a way that $2D\,$ locality
is restored \cite{BPZ}. In some cases this can be done in different ways
constrained by modular invariance of the WZNW partition function; for
$G=SU(2)\,$ this leads to the ADE classification of \cite{CIZ}.

The {\em canonical} quantization of the chirally split WZNW model
\cite{F1}-\cite{FHIOPT}
leads to a description in terms of chiral fields
revealing the quantum group (QG) invariance of their exchange algebras which
is the quantum counterpart of the Poisson-Lie invariance
of the underlying classical theory (see
\cite{BFP} for a recent comprehensive exposition of the classical situation and
\cite{GTT}, focused on the boundary WZNW model). The fact that the
monodromies of the chiral correlation
functions are related to $U_q({\cal G})\ 6j$-symbols (for $q\,$ an even root
of unity, $q=e^{\pm i \frac{\pi}{k+2}}\,$ for $G=SU(2)\,$)
has been known for a long time \cite{TK}.
On the other hand, it is clear that the true "internal" symmetry of the model is much
more involved (see e.g. \cite{VZ} and references therein for a recent
analysis of the relation between weak $C^*$-Hopf algebras and rational conformal
field theories). A plausible way out of this apparent contradiction
would be the assumption \cite{FHT3} that $U_q({\cal G})\,$ plays the role
of a generalized "gauge" group on the {\em extended} (chirally split) WZNW
so that one is facing an alternative analogous to choosing
unitary or covariant gauges in gauge theories, the latter necessarily including
unphysical states. In the case at hand this means that we have to consider
indecomposable representations of both the conformal current algebra
and of the quantum group.

For the $SU(2)_k\,$ WZNW model the minimal extension should involve primary fields
with isospins covering at least twice the range of the unitarizable representations,
$0\le 2I\le 2k+2\,,$ since the indecomposable QG counterparts relate $I\,$ with
$k+1-I\,.$ Allowing for non-integrable $I\,,$ one has to expect the extended
theory to be logarithmic \cite{RS}, \cite{Gu} (see e.g. \cite{KN} and
references therein). A fact, related to the latter, is that the commonly used
bases of KZ solutions \cite{ZF}, \cite{CF} are ill defined because some
mutual normalizations become inevitably infinite; fortunately,
(regular) bases are known \cite{STH} that remain meaningful in this wider
range of the isospins \cite{HST}. The elements of all these bases are given,
up to a prefactor which is an algebraic function of coordinate differences,
by multi-contour integrals in the complex domain
(identified, for $n$-point correlation functions, with the
Riemann sphere ${\Bbb C}P^1\,$ with $n-1$ punctures).
The braiding properties of the elements
of the regular bases have been also displayed in \cite{STH}, the corresponding
elementary braid matrices being triangular and well defined.

Let us consider the set of four-point conformal blocks of
$SU(2)_k\,$ WZNW chiral fields of isospin $I\,.$
Among them there is a distinguished set given by
{\em rational} functions.
For $0\le 2I\le k\,$ the importance of these
has been elucidated in \cite{MST}, \cite{ST},
\cite{RST} (see also \cite{ST2} where the more general problem of finding
all {\em algebraic} KZ solutions has been solved) -- they are in one-to-one
correspondence, provided that the conformal dimension
$\Delta_I = \frac{I(I+1)}{k+2}\,$ is an integer,
with the possible extensions of the corresponding chiral
algebra (the algebra of observables, in this case the current algebra).
In fact, only primary fields with integer or half-integer
conformal dimensions which are also local
with respect to themselves can have rational four-point functions.
Local commutativity (in the chiral sense) singles out the
$D_{even}\,$ series and the exceptional $E_6\,$ and $E_8\,$
models with diagonal pairing in the ADE classification.

The main objective of the present paper is to extend the
results of \cite{MST}, \cite{ST} finding rational solutions for non-integrable
values of $I\, (\, > \frac{k}{2} )\,$ as well.
After introducing our basic conventions
and notations in Section 2, in Section 3 we employ an alternative
description of the rational solutions of the KZ equation
as braid invariant linear combinations of the regular basis vectors. This
leads to nontrivial relations even in the known cases.
New rational solutions for $I = k+1\,$ are displayed in
the last Section 4 where we also analyze their properties. We hope to be able
to present an exhaustive study of this subject in the near future \cite{HP}.

\section{KZ equation and braiding properties of the regular basis}
\label{sec:2}
We will give here a short list of all needed notions and formulas;
for the lack of space we refer for details to \cite{STH}, \cite{MST},
\cite{ST} and \cite{HST}.

The conformal block containing four primary fields $\Phi_I (z)\,$
of isospin $I\,$ can be expressed as
$$
{\cal h} \,\Phi_I (z_1)\, \Phi_I (z_2)\, \Phi_I (z_3)\, \Phi_I (z_4)\, {\cal i}\, =\,
D_I (\underline{\zeta} ,\uz ) \, f_I( \x , \i )\,,
$$
\ba
D_I (\underline{\zeta} ,\uz ) &=& (\x_1 +\x_2)^{2I} \left(
\frac{ \i_1 +\i_2}{\i_1 \i_2} \right)^{2\Delta_I}\nonumber\\
&\equiv& (\zeta_{13}\zeta_{24})^{2I}
\left(\frac{z_{13}z_{24}}{z_{12}z_{34}z_{14}z_{23}}\right)^{2\Delta_I}
\lb{1}
\ea
(see \cite{ST}), where  $z_{ij}= z_i - z_j\,,\ \zeta_{ij} =\zeta_i - \zeta_j\,,$
\ba
&&\i_1 = z_{12}z_{34}\,,\quad  \i_2 = z_{14}z_{23}\,,\quad
\i =\frac{\i_1}{\i_1+\i_2}=\frac{z_{12}z_{34}}{z_{13}z_{24}}\,,\nonumber\\
&&\x_1 = \zeta_{12}\zeta_{34}\,, \quad \x_2 = \zeta_{14}\zeta_{23}\,,\quad
\x =\frac{\x_1}{\x_1+\x_2}=\frac{\zeta_{12}\zeta_{34}}{\zeta_{13}\zeta_{24}}\,.
\lb{2}
\ea
The function $f_I (\x ,\i )\,$ depending only on the harmonic ratios
is a {\em polynomial in} $\x\,$ of degree $2I\,;$
we are using the convenient polynomial bases of $SU(2)\,$ irreducible
representations $V_I\,$ and invariant tensors\footnote{ The relation of the latter
with the tensor invariants can be illustrated by the correspondence
$\x_1\ \leftrightarrow\ \e^{A_1 A_2}\e^{A_3 A_4}\,,\
\x_2\ \leftrightarrow\ \e^{A_1 A_4}\e^{A_2 A_3}\,$
in the simplest case $I=1/2\,$
when there are only two independent invariant
tensor structures. Here $\e^{AB}\,, \ A,B=1,2\,$ is the two dimensional
skew-symmetric tensor spanning ${\rm Inv}\, V_{1/2}^{\otimes 2}\,.$
In terms of the harmonic ratio the first invariant corresponds to $\x\,,$ and
the second -- to $1-\x\,.$}.

The corresponding KZ equation for $f_I (\x ,\i )\,$ reads
\ba
&&\left( (k+2) \i (1-\i ) \frac{\partial}{\partial\i} -
\sum_{i=0}^2  K^I_i (\x , \i )
\frac{\partial^i}{\partial\x^i}\right)\, f_I(\x ,\i ) = 0\,,\nonumber\\
&&K^I_0 (\x , \i ) = 2I \left( 2I (1 -\x ) - 2( I+1 )\i  +1 \right)\,,\nonumber\\
&&K^I_1 (\x , \i ) = (4I - 1 )\,\x^2+ 2\, \x \,(\i -2I ) -\i\,,
\nonumber\\
&&K^I_2 (\x , \i ) =  \x (1-\x )(\x -\i )\,.
\lb{3}
\ea
For any $I\,$ in the range $0\le 2I\le 2k+2\,$ Eq.(\ref{3}) has $2I+1\,$
(the dimension of ${\rm Inv}\, V_I^{\otimes 4}\,$) linearly
independent solutions.
We will define the regular basis vectors
$w_{I\mu} = w_{I\mu} (\underline{\zeta} ,\uz ) $, $\mu =0,\dots ,2I$
as in \cite{STH} (including the prefactor $D_I (\underline{\zeta} ,\uz )\,$)
\footnote{The conformal block (\ref{1})
has to be considered as a linear combination
of $w_{I\mu}\,$ with coefficients restricted further by
the locality condition imposed on the $2D\,$ correlation function.}.

As mentioned above, the solutions of the KZ equation (\ref{3}) are given
in terms of contour integrals defining, in general, multivalued analytic
functions of $\i\,,$ and we are interested in their {\em monodromy} properties.
In fact, the set of $w_{I\mu}\,$ is closed under {\em braiding}
("half monodromy") as well. In the case at hand (four equal isospins $I\,$)
the relevant braid group is ${\cal B}_3\,$ so that there are
two elementary braid
generators corresponding, respectively, to the transformations
\ba
B_1: &(&\x_1 ,\, \x_2 )\ \rightarrow \ (\, - \,\x_1 ,\,\x_1 + \x_2 )\,,
\nonumber\\
&(&\i_1 ,\, \i_2)\ \rightarrow \ (\, e^{-i\pi}\i_1 ,\, \i_1 + \i_2 )\,,\nonumber\\
B_2: &(&\x_1 ,\, \x_2 )\ \rightarrow \ (\, \x_1 + \x_2 ,\, -\,\x_2 )\,,
\nonumber\\
&(&\i_1 ,\, \i_2)\ \rightarrow \ (\, \i_1 + \i_2 ,\, e^{-i\pi} \i_2 )\,.
\lb{4}
\ea
Their action on the regular basis,
$$(B_i^{(k,I)} w )_{I\mu} = (B_i^{(k,I)})^\l_{~\mu} w_{I\l}\,,\quad i=1,2\,,$$
is given by
\ba
&&(B_1^{(k,I)})^\l_{~\mu}
= q^{2I(k+1-I)} (-1)^\l q^{\l (\mu +1)} \left[{\l\atop\mu}\right]\,,\nonumber\\
&&\l\,,\ \mu = 0,1,\dots , 2I\,,\ \ \left[{\l\atop\mu}\right] =
\frac{[\l ]!}{[\mu ]![\l - \mu ]!}\,,
\lb{5}\\
&&[\l ]! = [\l ] [\l -1]! \quad (\, [0]!=1\, )\,,\quad
[\l ] = \frac{q^\l - q^{-\l}}{q-q^{-1}}
\nonumber
\ea
(we will fix $q=e^{-i\frac{\pi}{k+2}}\,,$ see \cite{HST}) and
\be
B_2^{(k,I)} = F^I \, B_1^{(k,I)} \,F^I\,,\ ( F^I )^\l_{~\mu} = \d^{\l}_{2I-\mu}
\ \ ( (F^I)^2 = {1\!\!{\rm I}} )\,.
\lb{6}
\ee
The matrices $B_1^{(k,I)}\,,\ B_2^{(k,I)}$
are lower, respectively, upper triangular.

The rationality condition implies
that $f_I (\x ,\i )\,$ is a polynomial in $\i\,$ of order not exceeding
$4 \Delta_I\,$ \cite{MST}, \cite{ST}.
If $\Delta_I\,$ (and hence, $I\,$) is (half-)integer,
polynomial solutions of the KZ equation (\ref{3}) give rise to ${\cal B}_3\,$
invariants (up to a sign, for half-integer $\Delta_I$). Hence (see (\ref{1})),
\be
(-1)^{2I} f_I (1-\x , 1-\i ) = f_I (\x , \i ) =
\x^{2I} (- \i )^{4\Delta_I} f_I (\frac{1}{\x} ,
\frac{1}{\i} )\,.
\lb{7}
\ee

\section{Braid invariant functions in terms of the regular basis}
\label{sec:3}
All polynomial solutions of the KZ equation (\ref{3}) for $0\le 2I\le k\,$
(satisfying the initial condition $f_I (\x ,0) = \x^{2I}\,$ following from
the factorization of the four-point function into a product of two-point
functions for $\i\to 0\,$) have been found in \cite{MST}, \cite{ST}.
The list includes the simple currents series existing for any $k\,$
\be
f_{k/2} (\x , \i ) = (\x - \i )^k \quad (I=k/2\,,\ \Delta_{k/2} =k/4 )
\lb{8}
\ee
which, for integer $\Delta_{k/2}\,,$ corresponds to the $D_{even}\,$ series
in the ADE classification, and a few exceptional
cases occuring for $k=10\,,\ I=2,3\,$ and $k=28\,,\ I=5,9\,$
(corresponding to $E_6\,$ and $E_8\,,$  respectively;
see \cite{MST} for explicit expressions).
Except for the solutions in (\ref{8}) at odd values of $k\,,$ all the rest give
rise to rational functions. How could the latter be expressed in terms of the
multivalued functions of the 
regular basis?
To answer this question, one can make use of their simple braiding
properties. Taking into account the prefactors as well,
for all KZ solutions (\ref{8}) (including those for odd $k\,$)
one has
\ba
&&s^{(k, k/2)}= s^{(k, k/2)} (\underline{\zeta} ,\uz )
= D_{k/2} (\underline{\zeta} ,\uz )\, f_{k/2} (\x ,\i )\,,
\lb{9}\\
&&s^{(k, k/2)} = s^\mu w_{k/2~ \mu}\,,\quad
(B_{1,2}^{(k,k/2)})^\l_{~\mu} s^\mu = (-i)^k s^\l\,,\nonumber
\ea
i.e., $s^{(k, k/2)}\,$ are common eigenvectors of $B_{i}^{(k,k/2)}\,,\ i=1,2\,$
(see Eqs. (\ref{5}), (\ref{6}))
corresponding to the eigenvalue $(B_{1}^{(k,k/2)})^0_0 =
(B_{2}^{(k,k/2)})^k_k\,$ and hence the coefficients $s^\mu\,$ of their expansion in
terms of the regular basis can be found, up to an overall coefficient, by solving
a finite dimensional eigenvector problem. This can be easily done, and the solution
is (proportional to)
\be
s^\mu = \frac{(-1)^\mu}{[\mu +1]}\,,\quad \mu = 0,\dots , k
\lb{10}
\ee
-- one has to make use of a well known $q$-binomial identity
written in the form
\be
\sum_{\mu = 0}^k (-1)^\mu q^{\l (\mu +1)}\left[ {\l +1}\atop{\mu +1}\right] = 1
\quad {\rm for}\ 0\le \l\le k\,.
\lb{11}
\ee
For the ($E_6\,$) case $(k,I)=(10,3)\,$ one gets $s^0 = s^6 = 1,\ s^1 = s^5 =
-\frac{1}{[2]},\ s^2 = s^4 = \frac{1}{[3]},\ s^3 = \frac{3}{[3]}\,\frac{-1}{[4]}\,.$
(The apparent symmetry of the coefficients which is a general property can be
easily understood since $(s^\mu )\,$ should be an eigenvector of the antidiagonal
matrix $F\,$ (\ref{6}) as well \cite{ST2}.)

This result can be made more explicit in the (non-rational) case $k=1=2I\,$
for which the elements of the 
regular basis are known in terms of hypergeometric
functions \cite{HST} where it leads to the identity
\ba
&&(1-\i )^\frac{2}{3}\, (\, 2\, \x\, F_1 (1-\i )\, +
\, (1-\x )\, \i\, F_2 (1-\i )\, ) -\nonumber\\
&&- \,\i^\frac{2}{3}\, ( \,2\, (1-\x )\, F_1 (\i )\, +
\, \x\, (1-\i )\, F_2 (\i )\, ) =\nonumber\\
&&= \frac{2}{3}\, B(\frac{2}{3} ,\frac{2}{3} )\, (\x - \i )\,.
\lb{12}
\ea
Here $F_1 (\i ) = F (\frac{1}{3}, -\frac{1}{3}; \frac{2}{3};\i )\,,\
F_2 (\i ) = F (\frac{4}{3}, \frac{2}{3}; \frac{5}{3};\i )\,$ and $B(x,y)\,$ is the beta
function.

\section{Polynomial solutions of the KZ equation for I = k + 1}
\label{sec:4}
We have found polynomial solutions of the KZ equation (\ref{3}) for
$I=k+1=\Delta_{k+1}\,$ as well -- a value of $I\,$ corresponding
to a non-integrable representation of the current algebra,
the counterpart of the vacuum representation under the duality $I\ \leftrightarrow\
k+1-I\,.$ These solutions give rise to braid invariant vectors $(s^\mu )\,,\
\mu = 0,\dots , 2(k+1)\,$ whose only nonzero coefficient is $s^{k+1}\,$ 
and hence, in
contrast with the cases considered in the previous section,
the corresponding rational
functions belong to the regular basis. This fact might be important
since it can shed some light on the corresponding logarithmic CFT as well.

It is easy to obtain directly the polynomial solutions of
the KZ equation (\ref{3}) corresponding to $I=k+1\,$ for lower values of $k\,.$
Extrapolating the results, one arrives to the expression (throughout
this paragraph $I\equiv k+1\,$)
\ba
f_I(\x ,\i ) &=&(\i (1-\i ))^{I}
\sum_{m=0}^{2I}\sum_{n=0}^I C^I_{mn} \x^m \i^n \equiv\nonumber\\
&\equiv&
(\i (1-\i ))^{I}\, p_I (\x ,\i )
\lb{13}
\ea
where $p_{I} (\x ,\i )$ are polynomials of order $2I$ in $\x$ and of order $I\,$
in $\i\,,$ the coefficients $C^I_{mn}\,$ being chosen as
\be
C^I_{mn}=(-1)^{I+m+n}
{{I}\choose{m+n-I}} {{m+n}\choose{n}} {{3I-m-n}\choose{I-n}}\,.
\lb{14}
\ee
Note that the overall order of $f_I(\x ,\i )\,$ in $\i\,$ is $3I\,$ i.e.,
strongly below the upper bound $4\Delta_I\,.$
One can check directly that
the polynomial (\ref{13}) solves the KZ equation (\ref{3}) for $I=k+1\,.$
The expression for $p_I (\x ,\i )\,$ following from (\ref{13}), (\ref{14})
can be brought to the form
\be
p_I (\x , \i ) = \sum_{n=0}^I {{I}\choose{n}} \i^{I -n} \x^n P_{I n} (\x )
\lb{15}
\ee
where
\ba
&&P_{I n} (\x ) = \sum_{\ell =0}^I a_{n \ell}\, \x^\ell\,,\nonumber\\
&&a_{n \ell} = (-1)^\ell\frac{(I+\ell )!\,(2I-\ell )!}{\ell !\,(I-\ell )!(n+\ell)!
(2I-n-\ell)!}\,.
\lb{16}
\ea
Indeed,
\ba
&&p_I (\x , \i ) =
\sum_{n=0}^I \i^{I-n} \x^n \sum_{m=n}^{2I}(-1)^{m-n}\,\times
\lb{17}\\
&&\times\, {{I}\choose{m-n}} {{I+m-n}\choose{m}} {{2I-m+n}\choose{2I-m}}
\x^{m-n}\, =\nonumber\\
&&= \sum_{n=0}^I \i^{I-n} \x^n \sum_{\ell =0}^{2I-n}(-1)^\ell
{{I}\choose{\ell}} {{I+\ell}\choose{n+\ell}} {{2I-\ell}\choose{2I-n-\ell}} \x^\ell
\nonumber
\ea
which is equivalent to (\ref{16}). 
The form of the coefficients $a_{n \ell}\,$ (\ref{16}) implies that
\be
P_{I n}(\x ) = \frac{(-1)^n}{n!}\x^{2I+1}\frac{d^n}{d \x^n}
( \x^{n-2I-1} F(-I, I+1; n+1; \xi ) ).
\lb{18}
\ee
At these special integer values of the parameters the hypergeometric
functions in (\ref{18}) are expressible in terms of
Jacobi polynomials ${\cal P}^{(\a ,\b )}_\ell (x)\,$ \cite{BE},
\ba
&&{{n+I}\choose{I}}\, F(-I, I+1; n+1; \xi ) = {\cal P}_I^{(n,-n)} (1-2\xi )
\equiv\nonumber\\
&&\equiv \sum_{\ell =0}^I (-1)^{I-\ell} {{I+n}\choose{\ell}} {{I-n}\choose{I-\ell}}\,
\xi^{I-\ell} (1-\xi )^\ell =\nonumber\\
&&= \frac{1}{I!} \left( \frac{1-\xi}{\xi}\right)^n\,
\frac{d^I}{d \xi^I}\,\left( \xi^{I+n} (1-\xi )^{I-n} \right).
\lb{19}
\ea

It can be checked that $f_I (\x , \i )\,$ satisfies the two relations
of Eq. (\ref{7}). To prove this, one can use
the explicit expression (\ref{14}) for the coefficients 
$C^I_{mn}=(-1)^I\, C^I_{2I-m~I-n}\,$
and some combinatorial identities. Checking the second relation,
one should also have in mind that $f_I (\x , \i )\,$ only contains
powers of $\i\,$ in the range between $I\,$ and $3I\,.$

More details concerning the role of the corresponding rational
functions as vectors from the regular basis will be given elsewhere \cite{HP}.

The authors thank the organizers of the GIN 2001 conference
held in Bansko for their hospitality and financial support,
Prof. Ivan Todorov for his permanent interest
in this work and for valuable remarks, Yassen Stanev for discussion and
the Bulgarian National Council for Scientific
Research for partial support under contract F-828.

\end{document}